# Significant Interval and Frequent Pattern Discovery in Web Log Data

Dr. Kanak Saxena[1] and Mr. Rahul Shukla[2]

[1] Professor in Computer Application Department. R.G.P.V., S.A.T.I.
Vidisha, M.P., India

[2] M-Tech (Software System) Student, R.G.P.V., S.A.T.I.
Vidisha, M.P., India

**Abstract**
There is a considerable body of work on sequence mining of Web Log Data We are using One Pass frequent Episode discovery (or FED) algorithm, takes a different approach than the traditional apriori class of pattern detection algorithms.
In this approach significant intervals for each Website are computed first (independently) and these interval used for detecting frequent patterns/Episode and then the Analysis is performed on Significant Intervals and frequent patterns That can be used to forecast the user's behavior using previous trends and this can be also used for advertising purpose. This type of applications predicts the Website interest. In this approach, time-series data are folded over a periodicity (day, week, etc.) Which are used to form the Interval? Significant intervals are discovered from these time points that satisfy the criteria of minimum confidence and maximum interval length specified by the user.

## 1. Introduction

The expansion of the World Wide Web (Web for short) has resulted in a large amount of data that is now in general freely available for user access. The different types of data have to be managed and organized in such a way that they can be accessed by different users efficiently. Therefore, the application of data mining techniques on the Web is now focus of an increasing number of researchers. Several data mining methods are used to discover the hidden information in the Web. However, Web mining does not only mean applying data mining techniques to the data stored in the Web. The algorithms have to be modified Such that they better suit the demands of the Web. New approaches should be Used which better fit the properties of Web data. Furthermore, not only data mining algorithms, but also artificial intelligence information retrieval and natural language processing techniques can be used efficiently. Thus, Web mining has been developed into an autonomous research area.

Data Mining refers to the extracting or mining knowledge From large amounts of data' [3]. The growing interest in the field of data mining over the past few decades has resulted in a number of algorithms for processing various kinds of data including time-series data. In general, time-series data can be defined as an ordered sequence of values of a variable at specific (mostly periodic) time intervals. Time series data analysis is used in a variety of data-centric applications such as economic forecasting, sales forecasting, budgetary analysis, stock market analysis, inventory studies, census analysis and so forth. A considerable amount of work [6] has been done to process and mine through the large collections of time-series data sets using sequential mining. Most of the existing sequential mining techniques use individual time points, that is, events are considered to occur at specific time points. On the other hand, in many real-world scenarios, events are likely to occur with a high degree of certainty, not at specific time points, but within intervals. For example, it is useful to extract intervals of high activity from telephone logs to understand network usage.

In the Discovery of Frequent Pattern/Episode, it is important to predict website access in tight and accurate intervals to effectively predict the user behavior. Although there is a considerable body of work on sequential mining of transactional data, most of them deal with time-point data and make several passes over the entire data set in order to discover frequently occurring patterns We use an Approach in which significant intervals representing intrinsic nature of data are discovered in a single pass. In this approach, time-series data are folded over a periodicity (day, week, etc.) Which are used to form the Interval? Significant intervals are discovered from these time points that satisfy the criteria of minimum confidence and maximum interval length specified by the user. These Significant Interval are used to discover the frequent patterns in Web log data

In this Paper, we have used One PassSI Algorithm [4] and One Pass AllSI Algorithm [4] for generating the Significant Interval. These algorithm are single-pass algorithm, these algorithm use the main-memory approach to discover significant intervals in the Web Log Data. This approach not only makes use of a reduced data set by compressing point-based data to an interval-based data, but also makes only a single pass over the entire data set. We are also performing the analysis on One Pass SI algorithm [4], One Pass AllSI Algorithm [4] and One Pass FED algorithm [5].

The remainder of the paper is organized as follows. Section - 2 discussed the Related work , section-3 Defined the Related terms and Definition and Significant Interval.section-4 describe the process of significant Interval Discovery with the Example section-5 discuss the frequent pattern with the Example of Frequent Pattern Discovery and section-6 shows the Experimental Analysis on Significant Interval and Frequent pattern finally section-7 discussed the conclusion and future work

## 2. Related Work

The significant intervals used for the lock and unlock operations. These significant intervals can then be used to find an association or a relationship between the Websites. We have worked on the Dynamic Data (i.e. Web Log Data) that is dependent on the interest of user. These Intervals are used to discover the frequent pattern/Episode. Though the concept of intervals has been used for finding association rules ([11] Miller & Yang, 1997; [12] Srikant & Agrawal, 1996b), not many data mining algorithms discuss the formation of intervals on time-series data based on the interaction of events. WinEpi (Mannila et al., 1995) makes multiple passes over the data for counting the support of the candidates in each pass. Algorithm [4] is closer to MinEpi (Mannila et al., 1995) as we obtain the event count in a single pass over the data and use it to obtain support for the intervals. Regarding timing constraints, WinEpi and MinEpi find all sequences that satisfy the time constraint maximum span, which is defined as the maximum allowed time difference between latest and earliest occurrences of events in the entire sequence and minimum support, counted with the one occurrence per span window method. They have applied this approach on the static data but we are applying this approach on the Dynamic Data.

## 3. Terms and Definition and Significant Interval

We are Apply the One Pass SI algorithm[4],One Pass AllSI[4] and One Pass FED[5] on the web log data so we need the different website name such as citeseer.com,sports.com,newsworld.com etc The time-series can be represented with an Website timestamp model. A website w (for example citeseer.com is access, sports.com is not access, etc.) is associated with a sequence of timestamps {T1, T2, • ••, Tn} that describes its access over a period of time. The notion of periodicity (such as daily, weekly, monthly, etc.) is used to group the website accesses. For each website, the number of accesses at each time point can be obtained by grouping on the timestamp (or periodicity attribute). We term the number of accesses of each website as access count (ac). Thus the time series data can be represented as < w {Tl, al}, {T2, a2}, {T3, a3}, {Tn, an}>, where Ti represents the timestamp associated with the website w and ai represents number of accesses. ai can be referred as the access count of the website w at Timestamp Ti . This is referred to as folding of the time-series data using periodicity and time granularity (e.g., daily on seconds, daily on minutes, weekly on minutes).

### 3.1 Significant Interval

Intervals associated with a website are characterized by confidence and density. The confidence of a time point is the ratio (expressed as percentage) of the access count at that point to the periodicity of data collection (number of days or weeks (N)). When the interval consists of several time points, total access count of the interval (the sum of the access count of the points that form the interval) is used. The density of an interval is the ratio of the total access count of the interval to the interval length. We define a Window as a time-interval wd [Ts, Te] where (Ts <= Te), Ts is the start time and Te is the end time of the interval. An interval associated with a website is represented as [Ts, Te, ac, 1, d, c] where ac represents total access count of the interval, 1 denotes the length of an interval (Te-Ts+1), d indicates the density, and c represents confidence of the interval (ac/N * 100). Given a time sequence T, minimum confidence min-conf and interval length max-Len, we define the interval wd [Ts, Te] as a Significant Interval in T if:

1) Confidence(c) of wd >= min-conf
2) length (l) of wd <= max-Len
3) There is no window wd'= [Ts ', Te'] in T such that Ts '>=Ts and Te'=<Te that satisfies i) and ii).

### 3.2 Valid Significant Interval

Significant intervals can be of unit size, overlapping or disjoint. All valid intervals should be discovered by a significant interval discovery (SID) algorithm. Figure 3.1 shows all possible valid significant intervals. That B combines with C since B started after A. The Figure3.1 taken from [4].

1) Unit Significant Interval: It is a significant interval with the same start and end time; that is, $T_s = T_e$.

2) Disjoint Significant Intervals: They are defined as two significant intervals (that is, wd [$T_s$, $T_e$] and wd` [$T_s`$, $T_e`$]), which do not overlap (that is, $T_s` >= T_e$ and $T_e`$ is not in the interval [$T_s$, $T_e$] or $T_e`> T_s$ and $T_s`$ is not in the interval [$T_s$, $T_e$]).

3) Overlapping Significant Intervals: They are defined as two significant Intervals wd [$T_s$, $T_e$] and wd' [$T_s'$, $T_e'$]. (if $T_s<=T_e'<T_e$ and $T_s'<T_s$) or ($T_s<T_s'<=T_e$ and $T_e'>T_e$).

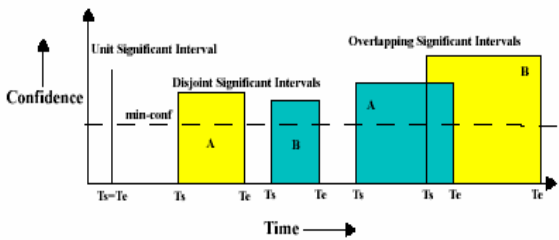

Figure3.1 Valid Significant Interval

### 3.3 Invalid Significant Interval

An invalid significant interval is identified by the third condition of the definition for a significant interval; that is, a significant interval cannot have an embedded valid significant interval (including a unit significant interval).

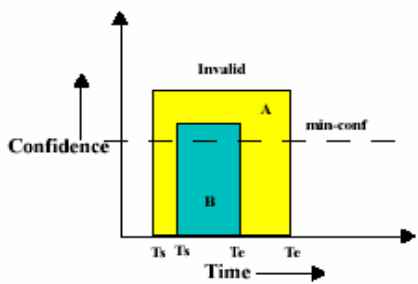

Figure3.2 Invalid Significant Interval

Figure 3.2 gives an example of an invalid significant interval. A is not a significant interval as it contains another significant interval B. Hence, only B is a significant interval. In other words, B is the tightest significant interval. The Figure3.2 taken from [4].

## 4. Process of Significant Interval Discovery

The Raw data is collected and then apply the preprocessing steps The Preprocessing steps are

1) Data Cleaning: In this step we are separate the data into the different website record and use the different views to show the record of each website. This process is called Data cleaning

2) Data Folding: In the preprocessing phase, the input data set is combined to periodicity of interest (example, daily or weekly). This process is called Data Folding.

Now One Pass SI or One Pass AllSI [4] is applied for generate the significant Interval. Those Time points satisfied the Minimum Constraint (i.e. min-conf and max Len for One Pass SI[4] or min-Conf for One Pass AllSI[4]) generate the Significant Interval are Discovered.

### 4.1 Example

Before applying One Pass SI or One Pass AllSI Algorithm [4] The Preprocessing steps must be taken.

Consider a 7 day data set. This dataset contains the Website name, its Access status (Access/Not Access), and timestamp (this combination is considered as a unique Website). There are two different website in the Table 4.1(raw data) so After Data Cleaning the Table 4.1 Break up in the two different Tables. Table 4.2 Data Cleaning (Citeseer.com) and Table 4.3 Date Cleaning (Rgtu.net).

Table 4.1 Raw data

| Website | Access Status | Timestamp |
|---|---|---|
| Citeseer.com | Access | 4/15/2009 2:05 pm |
| Rgtu.net | Access | 4/15/2009 2:10 pm |
| Citeseer.com | Access | 4/16/2009 2:10 pm |
| Citeseer.com | Access | 4/17/2009 2:40 pm |
| Citeseer.com | Access | 4/18/2009 2:40 pm |
| Rgtu.net | Access | 4/19/2009 2:10 pm |
| Citeseer.com | Access | 4/19/2009 2:05 pm |
| Rgtu.net | Access | 4/20/2009 2:20 pm |
| Rgtu.net | Access | 4/20/2009 2:10 pm |
| Rgtu.net | Access | 4/21/2009 2:05 pm |
| Citeseer.com | Access | 4/21/2009 2:05 pm |
| Citeseer.com | Access | 4/21/2009 2:10 pm |
| Rgtu.net | Access | 4/22/2009 2:05 pm |
| Citeseer.com | Access | 4/22/2009 2:10 pm |
| Rgtu.net | Access | 4/22/2009 2:20 pm |

Table 4.2 Data Cleaning (Citeseer.com)

| Website | Access Status | Timestamp |
|---|---|---|
| Citeseer.com | Access | 2/15/2009 2:05 pm |
| Citeseer.com | Access | 2/16/2009 2:10 pm |
| Citeseer.com | Access | 2/17/2009 2:40 pm |
| Citeseer.com | Access | 2/18/2009 2:40 pm |
| Citeseer.com | Access | 2/19/2009 2:05 pm |

| Citeseer.com | Access | 2/21/2009 2:05 pm |
| Citeseer.com | Access | 2/21/2009 2:10 pm |
| Citeseer.com | Access | 2/22/2009 2:10 pm |

Table 4.3 Data Cleaning (Rgtu.net)

| Website | Access Status | Timestamp |
|---|---|---|
| Rgtu.net | Access | 2/15/2009 2:10 pm |
| Rgtu.net | Access | 2/19/2009 2:10 pm |
| Rgtu.net | Access | 2/20/2009 2:20 pm |
| Rgtu.net | Access | 2/20/2009 2:10 pm |
| Rgtu.net | Access | 2/21/2009 2:05 pm |
| Rgtu.net | Access | 2/22/2009 2:05 pm |
| Rgtu.net | Access | 2/22/2009 2:20 pm |

Now the Data Folding step is taken on Table 4.2 (for citeseer.com) and table 4.3 (for rgtu.net).The result of Data Folding step is given in Table 4.4 and Table 4.5.The Table 4.4 and Table 4.5 contain the time point at which the website was 'Access' and number of times (over the entire dataset) it was 'Access' at that time point. For example, Citeseer.com has been `Access' at 2:05 three times (on three different days) in the data set (of seven days).

Table 4.4 Data Folding (Citeseer.com)

| Website | Access Status | Times of access | Access Count |
|---|---|---|---|
| Citeseer.com | Access | 2:05 | 3 |
| Citeseer.com | Access | 2:10 | 3 |
| Citeseer.com | Access | 2:40 | 2 |

Table 4.5 Data Folding (Rgtu.net

| Website | Access Status | Times of Access | Access Count |
|---|---|---|---|
| Rgtu.net | Access | 2:05 | 2 |
| Rgtu.net | Access | 2:10 | 3 |
| Rgtu.net | Access | 2:20 | 2 |

Consider a 7 day data set which gives a folded table as given in Figure 4.4 and 4.5. For 60% minimum confidence (min-conf), maximum interval length (max-Len) of 20 minutes and Daily periodicity, the One PassSI algorithm [4] performs the following steps for Table4.4:

1) The algorithm starts significant interval discovery at time point 2:05 and combines it with time point 2:10 and computes a confidence of 85.71% ((3+3)*100/7). Since the required 60% confidence has achieved by combining these two time points, and also satisfied the max-Len because the length is 5 minutes. Discovery of a significant interval (say A) with start time of 2:05 and end time of 2:10 and confidence of 85.71%.

2) Next, the algorithm starts significant interval discovery at time point 2:10 and combines it with time point 2:40 to give a confidence of 71.42 %(( 3+2)*100/7) with interval length of 30 minutes. Even though this interval satisfies the constraint of min-conf, it is not a significant interval because it does not satisfy the max-Len constraint.

3) Next, the algorithm starts significant interval discovery at time point 2:40 which cannot be further combined with any more time points. Since all time points in the data set have been considered there are no more time points and the algorithm stops. The significant Interval is given in Table 4.6

Table 4.6 Significant Interval Using One PassSI algorithm

| Website | Start Time | End Time | Interval Confidence |
|---|---|---|---|
| Citeseer.com | 2:05 | 2:10 | 85.71% |
| Rgtu.net | 2:05 | 2:10 | 71.42% |
| Rgtu.net | 2:10 | 2:20 | 71.42% |

Now for 60% minimum confidence (min-conf) in Daily Periodicity and max_Len is not required in the One Pass AllSI [4], the One Pass-AllSI algorithm [4] performs the following steps for Table4.4:

1) The algorithm starts significant interval discovery at time point 2:05 and combines it with time point 2:10 and computes a confidence of 85.71%. Since the required 60% confidence has achieved by combining these two time Points. Discovery of a significant interval (say A) with start time of 2:05 and end time of 2:10 and confidence of 85.71%.

2) Next, the algorithm starts significant interval discovery at time point 2:10 and combines it with time point 2:40 to give a confidence of 71.42% with interval length of 30 minutes. This interval satisfies the constraint of min-conf, it is a significant interval because there are No max_Len Constraint. As we see in previous, when we used One Pass-SI algorithm for significant interval discovery, even though this interval satisfied the constraint of min-conf, it was not discovered as a significant interval because it did not satisfy the constraint of max-Len. But here we are using the One Pass-AllSI [4] algorithm for significant interval discovery, this interval (say B) is classified as a significant interval with start time of 2:10, end time of 2:40 and confidence of 71.42%.

3) Next, the algorithm starts significant interval discovery at time point 2:40 which cannot be further combined with any more time points. Since all time points in the data set have been considered there are no more time points and the algorithm stops. The significant Interval is given in Table 4.7

Table 4.7 Significant Interval using One Pass AllSI Algorithm

| Website | Start Time | End Time | Interval Confidence |
|---|---|---|---|
| Citeseer.com | 2:05 | 2:10 | 85.71% |
| Citeseer.com | 2:10 | 2:40 | 71.42% |
| Rgtu.net | 2:05 | 2:10 | 71.42% |
| Rgtu.net | 2:10 | 2:20 | 71.42% |

## 5. Frequent Pattern/Episode

As we saw in the significant Interval Discovery confidence of an interval has been defined for a single website .However, One Pass FED algorithm [5] detects frequent episodes/pattern. Hence, we need to define the confidence of a pattern using the confidence associated with the websites forming the pattern. This is termed as Pattern Confidence (PC). The Pattern Confidence of an episode within an interval is defined as the minimum of the confidence of its constituent websites. When several websites access with different confidence in an interval, we can only infer that all websites access with minimum confidence in that interval. Hence, the confidence of a pattern interval represents the minimum number of occurrence of an episode within an interval. With frequently occurring patterns, Pattern Confidence underestimates the actual probability of the websites access together but retains its significance or order relative to the other patterns discovered. Pattern Confidence is used instead of access count in the One Pass FED algorithm [5]. Frequent pattern/episodes are discovered by combining/merging Website. The parameters that define how this combination/ merge take place are Sequential Window and Semantics. Sequential Window defines the maximum allowed time within which Website in a frequent pattern/episode may Start/End. The way in which this maximum allowed time is interpreted is defined by Semantics. There are two types of Semantics which a user may define: Semantics-s and Semantics-e. Semantics-s generates all possible combinations of websites, which Start within Sequential Window units of the first websites. Semantics-e, on the other hand, generates combinations of websites that Start and End within the Sequential Window units of the first websites. We have Developed the Frequent Pattern using Semantic-s and now we are working to generate the Frequent pattern Using Semantic-e.

5.1 Example

In this subsection, we shall explain the above One Pass FED Algorithm [5] with the help of an example. This will give a clear view of the algorithm. One Pass FED algorithm [5] works on the significant intervals generated by the significant interval discovery algorithms. For this example, let us assume that the significant intervals have been generated using one of the significant interval discovery algorithms and are given in Table5.0

Table 5.0 Significant Interval Discovered by SID

| Website | Start Time | End Time |
|---|---|---|
| Citeseer.com | 1:00 | 1:15 |
| Rgtu.net | 1:10 | 1:20 |
| Newsworld.com | 2:00 | 2:10 |
| Citeseer.com | 2:00 | 2:10 |
| Rgtu.net | 2:05 | 2:15 |

For this example, we shall consider Semantics-s and Sequential Window-30 minutes. The One Pass FED Algorithm [5] goes through the following steps:
The number of distinct Website (n) is found as 3, that is n = 3.The One Pass FED algorithm [5] starts with first significant interval [Citeseer.com, 1:00, 1: 15, 70], that is, start is pointing at this significant interval.
The next pointer moves to significant interval [Rgtu.net, 1:10, 1:20, 80]. The start of this significant interval lies within Sequential Window units (that is, 30 minutes). Hence, these two significant intervals form a 2nd level frequent pattern/episode [Citeseer.com, Rgtu.net 1:00, 1:20, 70]. The Pattern Confidence of the frequent episode is taken as the minimum of the two confidences which is 70 in this case. Now, the next pointer moves to significant interval [Newsworld.com, 2:00, 2: 10, 75]. The Sequential Window constraint is violated. Hence, the next pointer will not move any further. None of the significant intervals after this significant interval will not satisfy the Sequential Window constraint. Hence, no more frequent pattern can be further formed with [Citeseer.com, 1:00, 1:15, 70] as the start (base) significant interval. Now, the start pointer moves to the significant interval [Rgtu.net, 1:10, 1:20, 80]. The next pointer moves to significant interval [Newsworld.com, 2:00, 2:10, 75]. Again, the Sequential Window constraint is violated and no more frequent pattern can be discovered with the current base significant interval.
Now, the start pointer moves to the significant interval [Newsworld.com, 2:00, 2:10, 75]. The next pointer moves to significant interval [Citeseer.com, 2:00, 2: 10, 80]. Both these significant intervals combine to form a 2nd level frequent pattern/episode [Newsworld.com, Citeseer.com, 2:00, 2:10, 75]. The next pointer now moves to the significant interval [Rgtu.net, 2:05, 2:15, 70]. This combines with the current frequent pattern to form a 3rd level frequent pattern [Newsworld.com, Citeseer.com, Rgtu.net, 2:00, 2:15, and 70]. Now, as the value of n is 3, we can conclude that no more higher level frequent pattern will be formed using the base significant interval [Newsworld.com, 2:00, 2:10, 75] Now, the start pointer moves to the significant interval [Citeseer.com,2:00, 2:10,80] and the process will continue

Table 5.1 Second Level Frequent Pattern

| Second Level Frequent Pattern | | | | |
|---|---|---|---|---|
| Website1 | Website2 | Start Time | End Time | Confidence (%) |

| Citeseer.com | Rgtu.net | 1:00 | 1:20 | 70 |
| Newsworld.com | Citeseer.com | 2:00 | 2:10 | 75 |
| Citeseer.com | Rgtu.net | 2:00 | 2:15 | 70 |

Table 5.2 Third Level Frequent Pattern

| Third Level Frequent Pattern ||||||
|---|---|---|---|---|---|
| Website1 | Website2 | Website3 | Start Time | End Time | Confidence (%) |
| Newsworld.com | Citeseer.com | Rgtu.net | 2:00 | 2:15 | 70 |

Finally, the frequent episodes have detected and given in the Table 5.1 and 5.2.

## 4. Experimental Analysis

A number of experiments were carried out to compare the performance of One Pass SI by changing the confidence and max_Len and compare the performance of One Pass SI[4] and one pass AllSI[4] and also Check the No. of Frequent Pattern/Episode have been Discovered when we are changing the Sequential window length. The Website Interest is calculating in different months. The Algorithm One Pass SI [4], One Pass AllSI [4] and One Pass Fed Algorithm [5] were implemented in VB.Net as a Front End and SQL Server 2000 As the Back End. The data has been taken from the cyber café which is providing the services to their customer in the 24 hours.

Data set used for experiments Figure 6.1

| No of Days | No of Tuples | No of Websites |
|---|---|---|
| 90 | 1700 | 5 |

6.1 Effect of Varying Parameter on One Pass SI Algorithm

6.1.1 Effect of max_Len

In these experiment we want to test the effect of the changing the max-Len user parameter for the same min-conf and periodicity. The chart seen in Figure 6.2 shows the effect of varying the max-Len for a set of experiments keeping the min-conf = 40% and periodicity as Weekly.
As seen from the figure 5.2 as the max-Len increases the No. of Significant Interval intervals are also increases. This is because, as the max-Len increases the number of time Points which are checked for forming significant intervals Also increases.

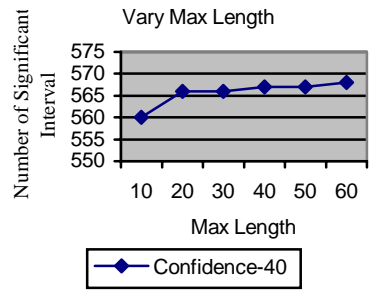

Figure 6.2

6.1.2 Effect of min-conf
Now we want to test the effect of the changing the min-conf user parameter for the same max-Len and periodicity. The chart seen in Figure 6.3 shows the effect of varying the min-conf for a set of experiments keeping the max-Len = 20 and periodicity as Weekly. As seen from the figure 5.3 as the min-conf increases the No. of Significant Interval intervals are decreases.

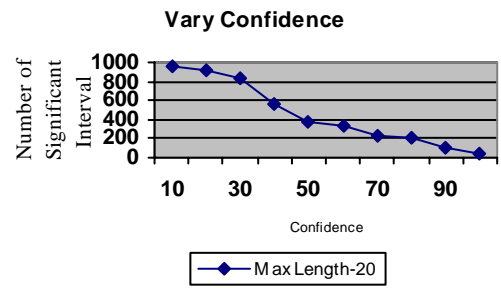

Figure 6.3

This is because as the min-conf increases the numbers of time points which are checked for forming significant intervals are decreases

6.2 Comparison of One Pass SI and One Pass AllSI Algorithm

6.2.1 Comparison between the numbers of Significant Intervals

An experiment was carried out to compare the number of significant intervals discovered by both the One Pass-SI and One Pass-AllSI algorithms [4]. The One Pass SI [4] has Discover more significant Interval than One Pass AllSI [4] up to confidence 60. This is because the One Pass-AllSI algorithm does not have a limitation on the max-Len but one Pass SI has this limitation When the confidence increases more than 60 One Pass SI [4] generates the more significant Interval as compare to One Pass AllSI [4] Actually this is dependent on the Data The parameter for Experiment was max_Len=20 minutes for One PassSI [4] comparison between the One Pass SI [4] and One Pass AllSI [4] is given in Figure 6.4.

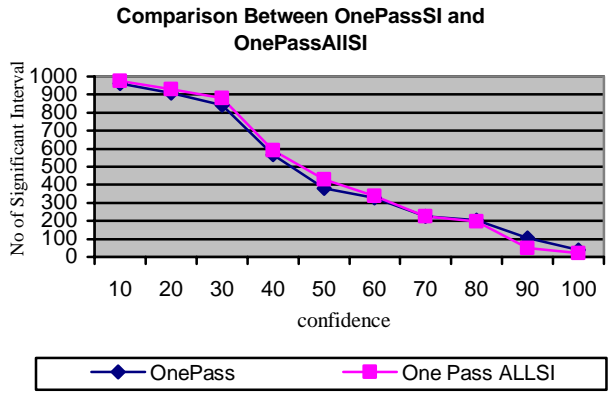

Figure 6.4

### 6.2.2 Time Comparison

As the One Pass AllSI algorithm [4] produces a superset of significant intervals as compared to its One Pass SI algorithm [4] counterpart, we wanted to check the extra time taken for finding all significant intervals at same min-conf. An experiment was carried out to analyze the difference in the time taken by each of these algorithms.
As seen from Figure 6.5, the One Pass-AllSI algorithm [4] takes more time than the One Pass-SI algorithm [4] for the different min-conf values.

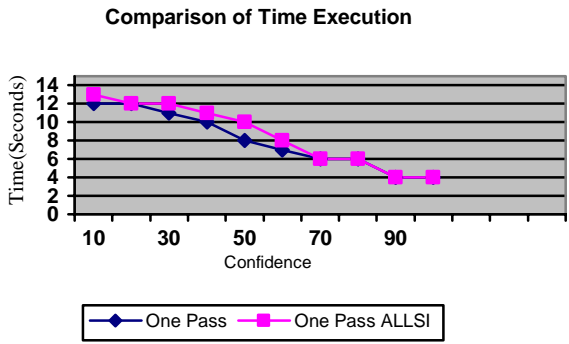

Figure 6.5

### 6.3 Effect of Varying Sequential Window Length on One Pass FED Algorithm

In the First phase of paper we have developed the significant Interval using One Pass SI algorithm [4] then we have applied the One Pass FED algorithm [5] for generate the Frequent Pattern/Episode. In this experiment we want to test the effect of the changing the sequential window length user parameter for the weekly periodicity. The Figure 6.6 shows the effect of varying the sequential window length. It is clear from the figure 6.6 when we increase the sequential window length the no. of FED also increase.

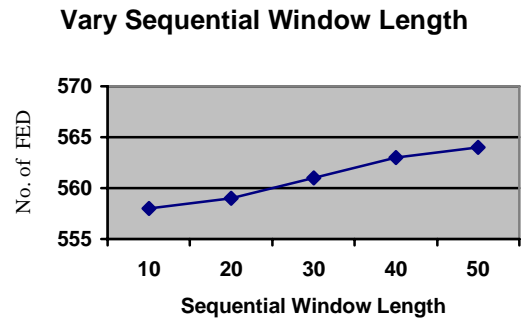

Figure 6.6

### 6.4 Contribution of Website in Frequent Pattern Discovery

In the final phase we are showing the interest of user in the particular website in the particular period of time. We have taken the data of month April, May and June. In the figure 6.7 we are showing the contribution of website in the Frequent pattern/Episode Discovery. The table-a gives the full form of website which are use in the Figure6.7.

| Table-a | |
|---|---|
| C | Citeseer.com |
| NC | Newsworld.com |
| SC | Sports.com |
| RN | Rgtu.net |
| EC | Election.com |

In the starting of April the citeseer.com, newsworld.com sports.com, Election.com is heavily accessed because this is the time of Election and IPL Tournament and M-Tech research student are also doing their research. In the month of May the people are also interested in the Rgtu.net because this is the time of admission in the different courses and this time result of different courses are also declared. In the month of June interest in Rgtu.net little bit increase as compare to previous month.

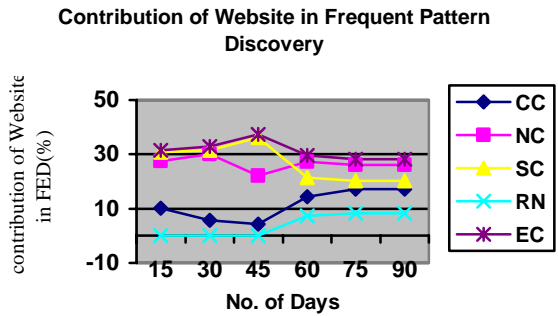

Figure 6.7

## 7. Conclusion and Future Work

In this paper, we have presented the discovery of significant intervals in Web Log Data Using One Pass SI and One Pass AllSI algorithm [4] and these significant Intervals are used to generate the Frequent Pattern. The frequent Pattern is generating by the One Pass FED algorithm [5]. These patterns are used to forecast the trend. We are applied this [4] for web Log Data Which is the Dynamic nature Data. And analysis performs on Web Log Data. And Experimental result shows the result of this analysis. Currently, we are working on Frequent Pattern Generation using Semantic-e on time-series web Log Data This process applied for other domain, such as traffic analysis, and others.

Dr. Kanak Saxena is the Professor in the Computer Application Department of Samrat Ashok Technological Institute, Vidisha (M.P.), India. This Institute is Affiliated Rajiv Gandhi Proudyogiki Vishwavidyalaya, Bhopal (M.P.), India

Mr.Rahul Shukla is the Student of M-Tech (Software System) from Samrat Ashok Technological Institute, Vidisha (M.P.), India